\documentclass[twocolumn]{article}

\usepackage{savesym}
\usepackage{stmaryrd}
\usepackage{amsfonts}
\usepackage{pifont}
\usepackage{bbding}
\usepackage{amssymb}
\usepackage{amsmath, amsthm, amssymb, upgreek}
\usepackage{graphicx, subfigure}
\usepackage{calrsfs}
\usepackage{wasysym}
\usepackage{xcolor}
\usepackage{extarrows}
\usepackage{mathrsfs}
\usepackage{mathtools}
\usepackage{subfigure}
\usepackage{float}
\usepackage{array}
\usepackage[none]{hyphenat}
\usepackage{url}

\title{WristAuthen: A Dynamic Time Wrapping Approach for User Authentication by
Hand-Interaction through Wrist-Worn Devices}

\author{
 Qi Lyu\footnote{\tiny School of Mathematics and Statistics, Xi'an Jiaotong University}
 \quad Zhifeng Kong$^*$ 
 \quad Chao Shen\footnote{\tiny School of Electronic and Information Engineering, Xi'an Jiaotong University}
 \quad Tianwei Yue$^*$
 }
\date{}

\begin{document}

\maketitle

\begin{abstract}
The growing trend of using wearable devices for context-aware computing and pervasive sensing systems has raised its potentials for quick and reliable authentication techniques. Since personal writing habitats differ from each other, it is possible to realize user authentication through writing. This is of great significance as sensible information is easily collected by these devices. This paper presents a novel user authentication system through wrist-worn devices by analyzing the interaction behavior with users, which is both accurate and efficient for future usage. The key feature of our approach lies in using much more effective Savitzky-Golay filter and Dynamic Time Wraping method to obtain fine-grained writing metrics for user authentication. These new metrics are relatively unique from person to person and independent of the computing platform. Analyses are conducted on the wristband-interaction data collected from 50 users with diversity in gender, age, and height. Extensive experimental results show that the proposed approach can identify users in a timely and accurate manner, with a false-negative rate of 1.78\%, false-positive rate of 6.7\%, and Area Under ROC Curve of 0.983 . Additional examination on robustness to various mimic attacks, tolerance to training data, and comparisons to further analyze the applicability.
\end{abstract}


\section{Introduction}

Nowadays the authentication and identification process of mobile devices are becoming increasingly important. For instance, many people are using online banking, and their property are being threatened directly by potential attacks.
In general, there are three types
of authentication factors: $(i)$ a password or a PIN code;
$(ii)$ some substance such as a security token; $(iii)$ the biometrics feature, such as fingerprints. The password is the usual
way we use, but \cite{Password1,Password2,Password3}
have shown that they have some security problems.
The identification by biometrics features developed nowadays seems
to provide a more accurate and safe approach for authentication.
There are also two types of methods by biometrics traits: using
physiological features such as fingerprint or iris \cite{Fingerprint1,Iris1}, or using behavioral features such as gait or keystroke. Moreover, attackers might
even get users' faces and fingerprints from public events and then use these biometrics for
authentication \cite{Attack1}. In contrast, some behavior-based
methods can be deal with attackers better because it is more difficult to
be simulated or forged.

\subsection{Wrist-worn Devices for Activity Recognition}
Wrist sensing, with data collected from wrist-worn devices, has many applications.
\cite{WristSense1} discovered
that wrist activity can help distinguish sleep from
wakefulness. They distinguished in
approximately $88\%$ of the time. \cite{WristSense2} developed a usage of a novel
usage of wrist motion: They used a watch-like sensors to continuously
track wrist motion throughout the day and detect periods of
eating. They described an algorithm that segments
and classifies eating periods and finally obtained an accuracy of $81\%$ for 1-s resolution.

In the authentication field, some researchers \cite{Biometrie} found the photoplethysmographic (PPG) sensor in smart-watches, which is usually used for catching heart rate, could distinguish a specified hand motion done by the user. Their experiments shown that a continuous wrist movement of three times achieved an average error rate of 11.6$\%$, and a movement of nine times achieved an average error rate of 8.8$\%$. However, the data collected by PPG sensor might be affected by user's physical condition in that a person's heart rate can be much higher after strenuous exercise.

\subsection{Wrist-Worn Devices for Behavioral Authentication}
Recently authentication by wrist-worn devices
come out with various approaches. \cite{Wrist1} developed a
motion-based authentication for wrist worn devices, by
histogram method and dynamic time warping method. They
set some gesture for authentication. The EER value could be as low as
$2.6\%.$ However, compared to our method, the range of the hand motion is too large (for example, draw a huge circle in front of user's body), which is not convenient in public
places. What's more, it may be forged and attacked by attackers, which
is not as safe as our method.

\cite{Wrist2} used digital wristband to distinguish user's
behavior after a user logged in. They analyzed user's
habit of using mouse and keyboard, and then deauthenticate the current user or not. They could verify $85\%$ of users in 11 seconds and $90\%$ in 50 seconds. However, their method mainly
solve the risk that when the user forget to log out after using.
Actually, they have to spend 50 seconds to get $90\%$ correct rate,
with the system running background. In contrast, our method focuses on
instantaneous pass of authentication.

\cite{Signature} worked on signature verification by wrist-worn devices. They successfully determined whether the signature is genuine or forged. They collected data with the accelerometer and gyroscope, then extracted features with dynamic time warping (DTW) and trained some classifiers. Finally they obtained 0.98 AUC and 0.05 EER. Despite the high accuracy, there are some disadvantages in comparison with our method. They require the labels of forged and genuine signatures to be given. However, we don't have forged signature data in practice. Because limited forged signature data, the system can't cover all situations. This will result in an unexpected classification for other input not included in the training data. This flaw will be illustrated in detail in Section \ref{SVM}. We only needs genuine signatures and can distinguish forged signatures automatically.

\subsection{Challenges and Contributions}
In general, people's writing behaviors vary greatly
from one to another,
but there are still several challenges. $(i)$ Wrist movement data is hard to distinguish with small letters, while writing larger
words produces more distinguishable data. In order to increase the usability and capability to defend simulated attacks, we restrict the minimum size of every letter with a square with side length $2cm$. $(ii)$ Data caught by accelerometer and gyroscope can be different when the same one writes the same word, since they might move quicker in one letter and slower in another, or make unexpected stop somewhere. So we chose DTW(Dynamic Time Warping) algorithm to cover these complex situations, enabling the similarities in writing stand out of the discrepancies. $(iii)$ There is no direct relationship between the path of wrist and writing information since the former one does not imply finger movement. As a result, it is not effective if we simply calculate wrist path from wrist movement directly for authentication.

Following are some advantages
of our method: $(1)$ Hard to attack. If
someone knows the word, the font is hard to acquire. Even if the
attacker knows how the user writes the word, it is difficult to simulate the writing habit to forge and through authentication. Related experiments will demonstrate
these. $(2)$ Efficient and convenient. It does not
need much calculation and can be finished in 1 to 2 seconds. $(3)$ Fitting in with office or class, where people usually use a pen.
$(4)$ Suitable for protecting some extremely important
things, such as safety box, classified documents and so on. Our
method can also provide a secondary password in situations like online paying.

\section{Writing Characterization}
\subsection{Writing Sensing}
Among all available sensors \cite{SDK}, we selected the accelerometer and the gyroscope, which could capture wrist movements precisely. Our method is not restricted by the type of digital devices, since these two sensors exist in most of the wrist-worn devices nowadays.
The accelerometer measures the acceleration of the Band in $X$, $Y$ and $Z$ axes. A value of 1 means the band is under $1g$ of acceleration, where $1g = 9.81 (m/s^2)$. The gyroscope measures the angular velocity of the Band also in $X$, $Y$ and $Z$ axes. When the angular velocity is 1 degree per second $(^\circ/s)$, the value will be 1.

\subsection{Data Collection}
Fifty distinct individuals' wrist movement data are collected with 62 times per-second sample rate. Each individual is required to write (using their right hands) the word "love" and an arbitrary word with 3 to 5 letters in lowercase, at least 6 times for each word. In order to capture personal characters in writing precisely, they are required to keep the writing speed in their normal level and control the size of each word to be larger than $10 \times 3~cm^2$.

The raw data is recorded as a tuple in sequence of time, with 6 motion dimensions in each tuple $v(t)=(a_x(t),~a_y(t),~a_z(t),~\omega_x(t),~\omega_y(t),~\omega_z(t))$, which denote the acceleration and angular velocity along axis $X,~Y,~Z$ and $t$ denotes time. Finally we collected 600 samples in total, while each file describing one person's wrist movement of a writing particular word in 6 motion dimensions.

\begin{figure}[htbp]
\centering
\includegraphics[width=2.3in]{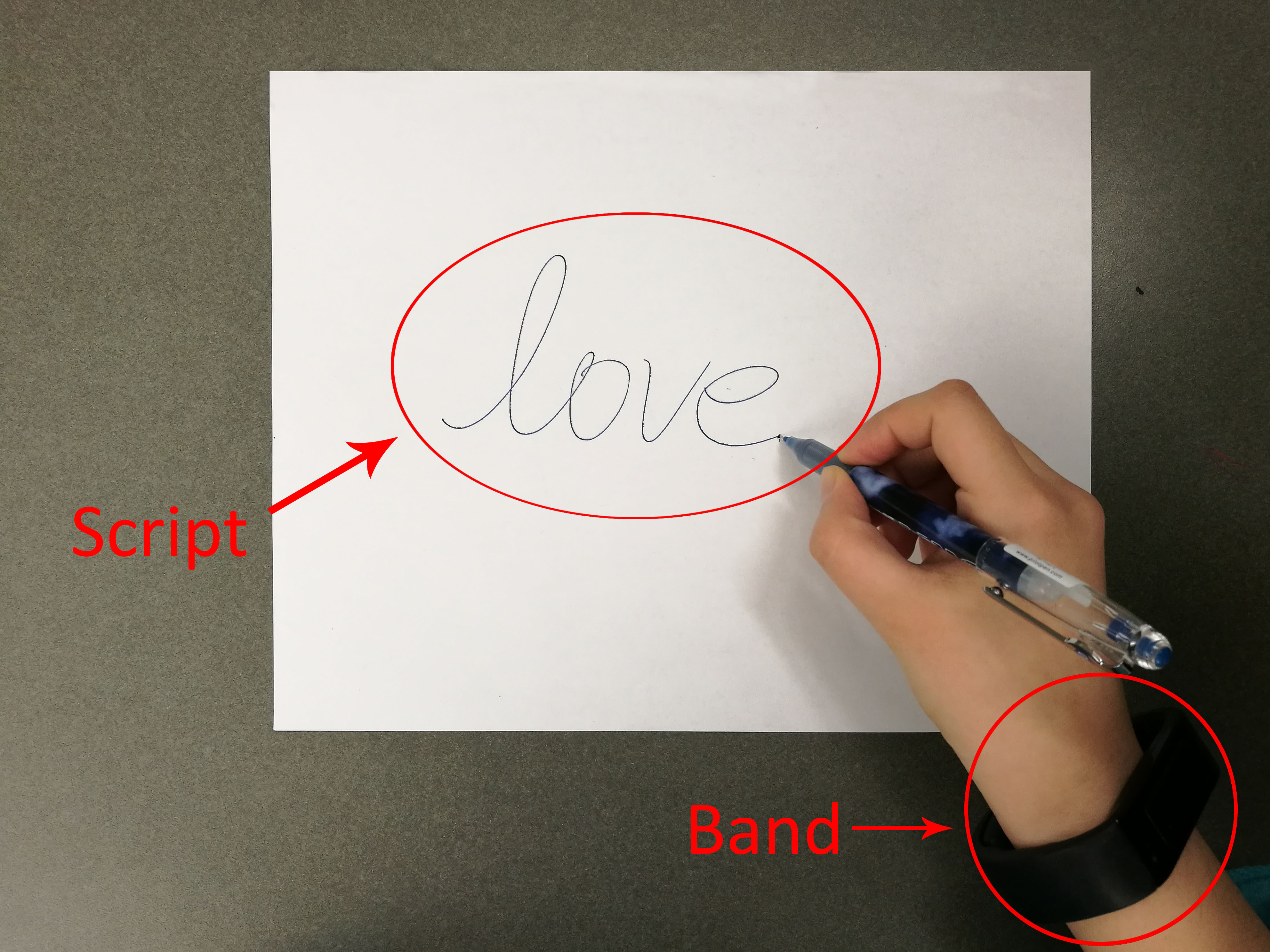}
\caption{Raw writing data is collected.} \label{fig3_1_5}
\end{figure}

\begin{figure}[htbp]
\centering
\includegraphics[width=3.3in]{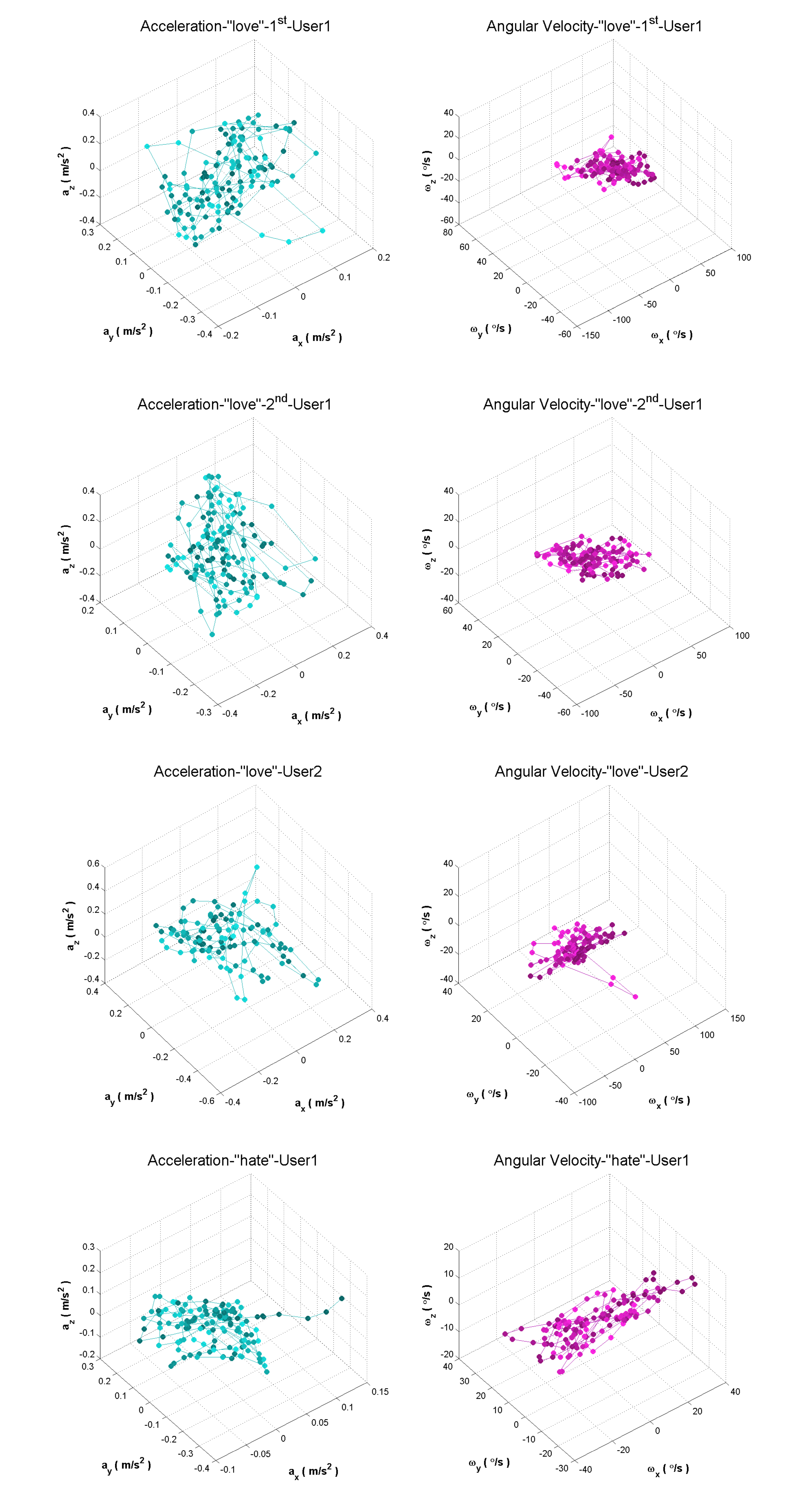}
\caption{Examples of the test data.} \label{fig3_1_alldata}
\end{figure}

The scenario of collecting raw data is shown in Fig.~\ref{fig3_1_5}. Examples of the test data we collected are shown in Fig.\ref{fig3_1_alldata}. The time order is expressed by the shading of the color, where light color points are before deep points.

\subsection{Data Choosing and Filtering} \label{Filtering}
\begin{figure}[htbp]
\centering
\includegraphics[width=3.0in]{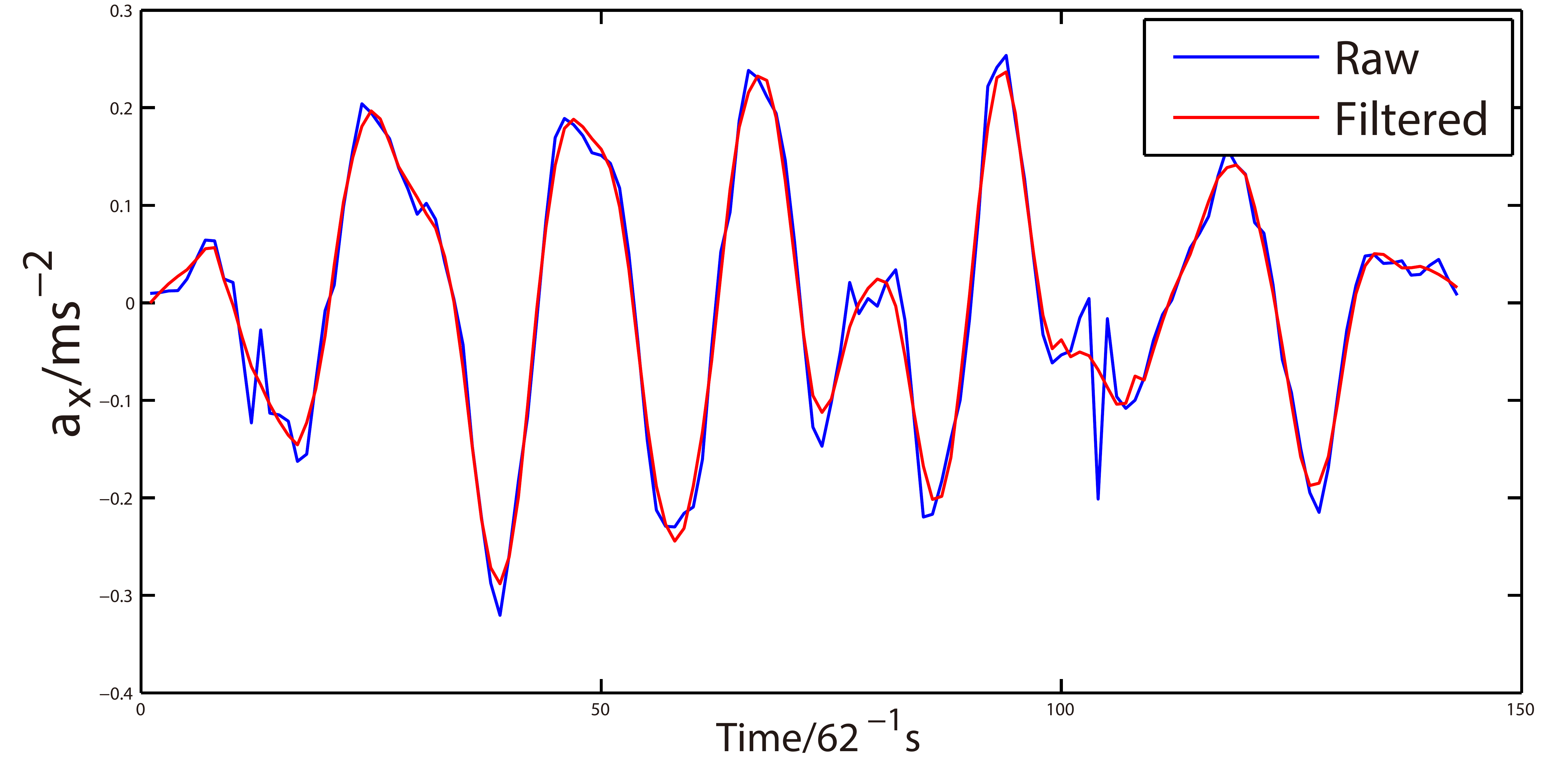}
\caption{Raw data and filtered data.} \label{fig3_3_3}
\end{figure}

It is hard to extract features in such high dimension, so we deal with $a_x,a_y,a_z,\omega_x,\omega_y,\omega_z$ versus time respectively in 2-D. For signal denoising and smoothing, we adopted Savitzky-Golay filter \cite{SGfilter}, a digital filter that increases the signal-to-noise ratio while keeping the features of signal, to preprocess our data. This is done by fitting successive nearby data points with a low-degree polynomial. To find the polynomial, we equally spaced the points and constructed least-squares equations for solution. In this paper, we chose each set of 9 successive points for smoothing and the degree of polynomial is 2.
We had every point filtered except for first four and last
four points, since there is not enough points around them. For those points, we reduced the length of the successive points set and
did similar process.
The denoising effect is shown in Fig.~\ref{fig3_3_3}. The main shape
of the curve does not change while the curve gets smoother.
Extremely peak values in a small neighborhood region are also
corrected by S-G filter.

\subsection{Distance Measurement}
It is necessary to find a robust and efficient algorithm to measure the distance between two time series. Actually, two time series share very similar shapes but not aligned in the time axis. Therefore, the Euclidean distance or $L-1$ norm fails to capture the similarity between them.

\begin{figure}[htbp]
\centering
\includegraphics[width=3.3in]{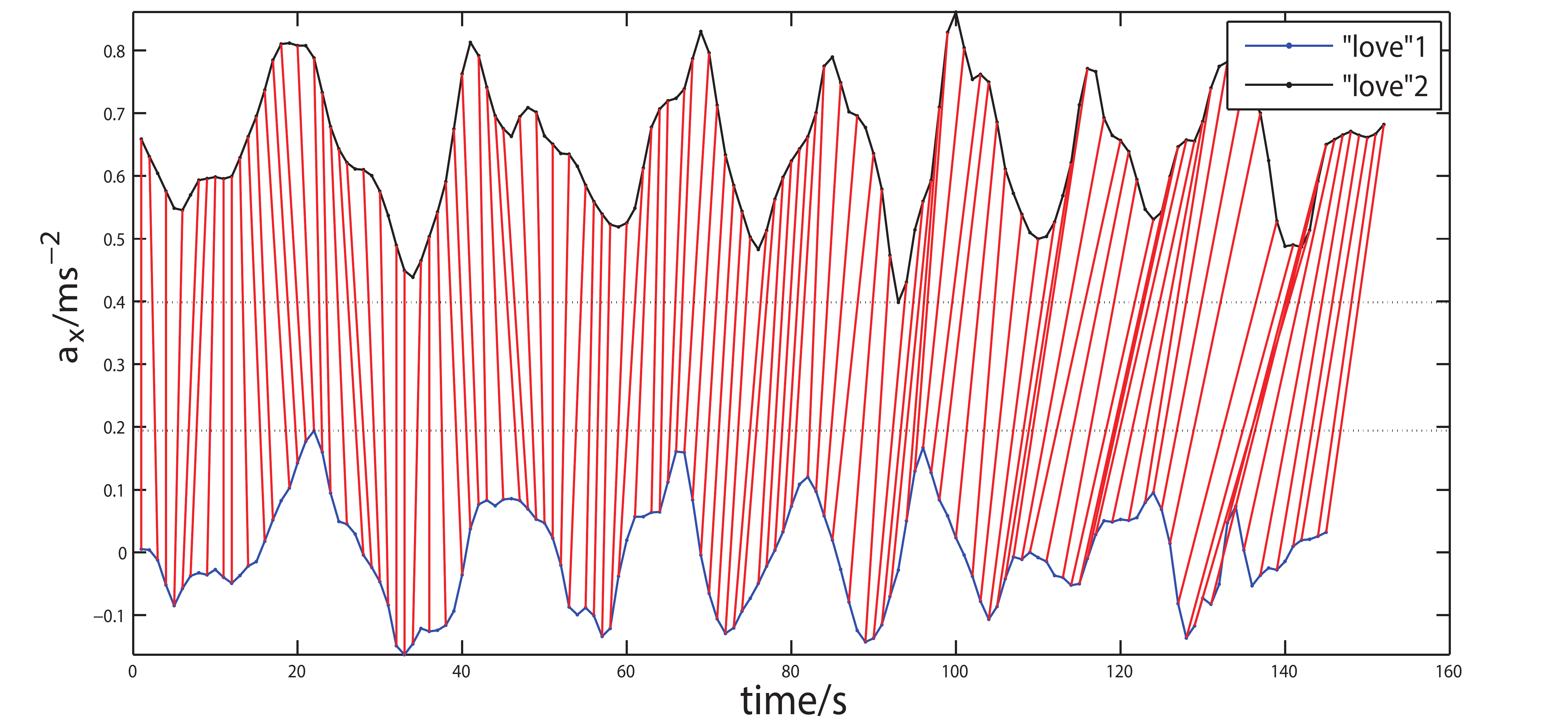}
\caption{The DTW bridge.} \label{fig3_4_2}
\end{figure}

The dynamic time warping method (DTW) meets our requirement \cite{DTWpaper}. It provides a more intuitive distance measurement, which is shown in Fig.~\ref{fig3_4_2}. It matches two time series with similar shapes, even if they are not closed under Euclidean distance. Below is a short explanation for DTW algorithm.

Assuming there are two time series $A$ and $B$ of length $m$ and
$n$, where
\[
A=(a_1,a_2,\cdots,a_m);\
B=(b_1,b_2,\cdots,b_n).
\]
Then a $m$-by-$n$ matrix $d$ is built, where the $(i^{th},j^{th})$
element of $d$ refers to distance between $a_i$ and $b_j$:
$(a_i-b_j)^2$ . Now we want to find a route from $(1,1)$ to $(m,n)$.
Note the route as $R$, where $R(k)$ is the $k^{th}$ points we stand
on. The length of $R$ is noted as $K$. Then we have
\begin{align*}
&\max(m,n)\leq K<m+n-1;\\
&R(1)=(1,1),~~R(K)=(m,n);\\
&R(t+1)-R(t)\in\{(0,1),(1,0),(1,1)\}.
\end{align*}
And we are interested in the path reaching minimum:
\begin{align*}
DTW(A,B)=\min\limits_R\left\{\left(\sum_{k=1}^Kd(R(k))\right)^{\frac{1}{2}}\right\}.
\end{align*}
The path is solved by dynamic programming, with Machine Learning Toolbox created by \cite{DTWtoolbox}. Define the cumulative
distance $s(i,j)$ and the distance $r(i,j)$ which have been found in
the current cell. We have
\begin{align*}
s(i,j)=r(i,j)+\min\{s(i-1,j-1),s(i-1,j),s(i,j-1)\}.
\end{align*}
DTW has very low computational complexity $O(mn)$.

\section{Authentication Architecture}
\subsection{Overview}
\begin{figure}[htbp]
\centering
\includegraphics[width=0.5\textwidth]{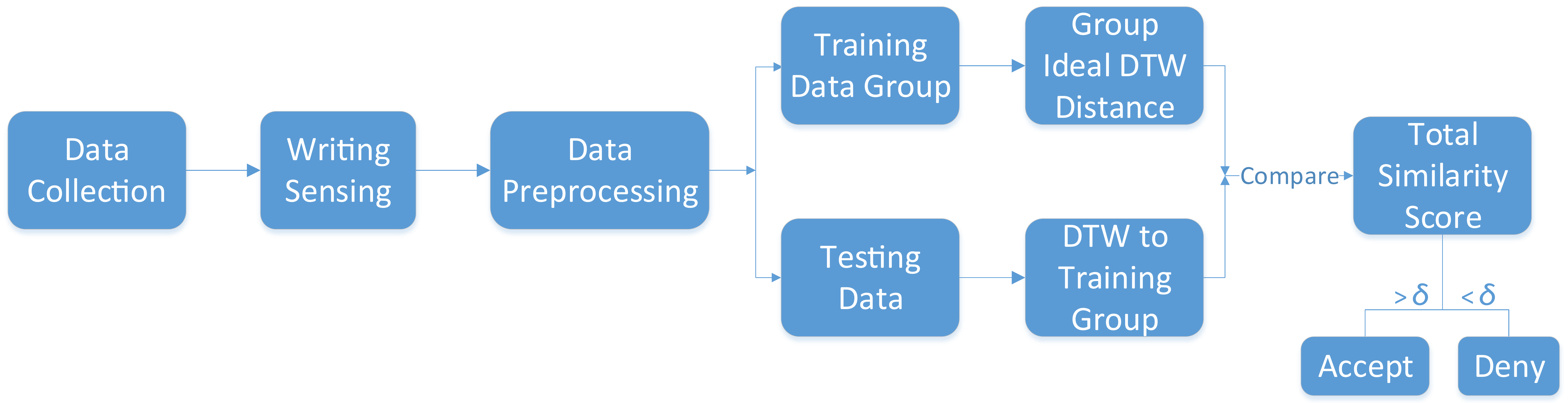}
\caption{The Flow chat of our system.}
\label{authentication_process}
\end{figure}

The authentication process consists of 4 parts: sensing, filter, trainer and identifier. The flow chat is shown in Fig~\ref{authentication_process}.
\subsubsection{Sensing}
We used data collected from the accelerometer and the gyroscope of wrist movement to express the writing behavior.
\subsubsection{Filter}
We used the S-G filter introduced in Section~\ref{Filtering}. The original data was smoothed and the points whose absolute values are extremely large were modified .
\subsubsection{Trainer}
The DTW distance was selected to do the authentication. To train our system, we obtained a group of data by letting the user write the same word for several times, and then we calculated the ideal DTW distance of this group of data. This concept is introduced in the next section.
\subsubsection{Identifier}
For a given testing data, we calculate the DTW distance from the testing data to the group of training data at first, then compare it with the ideal DTW distance of the group. Finally the total similarity score of the testing data is calculated to decide whether to accept or to deny.

\subsection{Group Characteristic}
A proper measurement is needed to capture the similarity within in one person's writing style while highlight the discrepancies among people. Actually we only have one class of data, in that the user sets the "password" by writing a certain word for several times. So we put forward two definitions here:
\subsubsection{Group ideal DTW distance}
The word "ideal" means our expectation of testing data for passing the authentication. If the user produces a testing data whose DTW distance to the group is closed to the ideal DTW distance, then the user could pass the authentication. For a group of training data $D_1,D_2,\cdots,D_n$, calculate DTW distance $\mathbf{d}_{ij}~(1\leq i\leq n,~i<j\leq n)$ between $D_i$ and $D_j$, where $\mathbf{d}_{ij}=(d_{ij}^{(1)},d_{ij}^{(2)},\cdots,d_{ij}^{(6)}),$ corresponding to 6 motion dimensions $a_x,a_y,a_z,\omega_x,\omega_y,\omega_z$. Then for each $k$ from 1 to 6, choose $\{d_{ij}^{(k)}\}$'s upper quartile $e_{ij}^{(k)}$ for ideal distance. From such calculation we get group ideal DTW distance $\mathbf{e}=(e^{(1)},e^{(2)},\cdots,e^{(6)})$.
\subsubsection{DTW distance to a group}
For testing data $T$, we intend to measure its distance to the training group $D$. During our tests, we found the speed of writing and the size of the word were different each time, although the user was asked to keep the speed and size. So we calculated the final DTW distance by proper weights. We calculated distances by 6 motion dimensions and got $\{t_i^{(k)}\}$, which means the DTW distance between $T$ and $D_i$ of $k^{th}$ motion dimension. Then we got $\{t_i^{(k)}\}$ sorted and obtained $\{\hat{t}_i^{(k)}\}$, where $\hat{t}_i^{(k)}\leq\hat{t}_j^{(k)},~if~i<j$. After that we chose PDF of Poisson Distribution for weight:
\[
\beta_i=\frac{\lambda^i}{i!}e^{-\lambda},~\lambda=\left[\frac{n}{5}\right],~\rho_i=\frac{\beta_i}{\sum_{j=1}^n\beta_j},~1\leq i\leq n.
\]
Then Final DTW distance is $\mathbf{s}=(s^{(1)},s^{(2)},\cdots,s^{(6)}),$ where
\begin{align*}
s^{(k)}&=\sum_{i=1}^n\rho_i\hat{t}_i^{(k)}.
\end{align*}

By feature of Poisson Distribution, the weights mainly distribute
near $[n/5]$, and for those distances ranked over $[n/2]$, the
weights are almost 0. That is, for a given testing data, the more similar training data is given higher weight when we calculate the DTW distance from the testing data to the training group.

As the result, the most outstanding advantage is to have a high fault tolerance of input training data. That is, if there are some bad training data mixed in the training set, the system can automatically ignore those bad data. The performance of the fault torlerance will be discussed by experiments later in section \ref{FaultTolerance}.

\subsection{Similarity Measurement and Authentication}
The judgment happens when a user wants to pass the authentication. First, the user wears the Band and writes a word as the testing data. Then the system calculates the total similarity score between it and training data. If the score is high enough, the access query will be accepted, otherwise denied.

Based on these definitions of group, when a testing data comes into the system, we easily define the similarity score $\mathbf{SS}=(SS^{(1)},SS^{(2)},\cdots,SS^{(6)})$ by 6 motion dimensions:
\begin{align*}
SS^{(k)}=\max\left\{\frac{e^{(k)}}{s^{(k)}},1\right\},~k=1,2,\cdots,6.
\end{align*}
Set weight $\boldsymbol{\mu}=(\mu^{(1)},\mu^{(2)},\cdots,\mu^{(6)})$
for 6 motion dimensions, where $\sum_{i=1}^6\mu^{(i)}=1$. Then the $TSS$ (Total Similarity Score) is
\[TSS=\sum_{k=1}^6\mu^{(k)}SS^{(k)}.\]
We can simply set $\mu^{(1)}=\mu^{(2)}=\cdots=\mu^{(6)}=1/6$. We
did AUC test and got better $\boldsymbol{\mu}$ for
each specific system. At last we set a threshold $\delta$. If
$TSS\geq\delta$, the user passes the authentication, otherwise
denied.

Another important issue is to complete the authentication process
quickly. The Computational Complexity of our system is $O(Ltn^2)$
since we can store the ideal DTW distance into the system,
where $L$ denotes the motion dimensions,
$t$ denotes the number of sampling points,
and $n$ denotes the number of training groups.

\section{Experiments and Evaluation}

In this section several experiments are put forward to evaluate the performance of our system. We demonstrate that our system can distinguish well between the authorized users and different types of mimic attackers. The system is also well designed to tolerate some improper input data in the training data set, and the performance of the system can be improved if we use the personalized signatures or patterns as the password.

\subsection{Self and Non-Self Discrimination}\label{4.1}
\begin{figure}[htbp]
\centering
\includegraphics[width=2in]{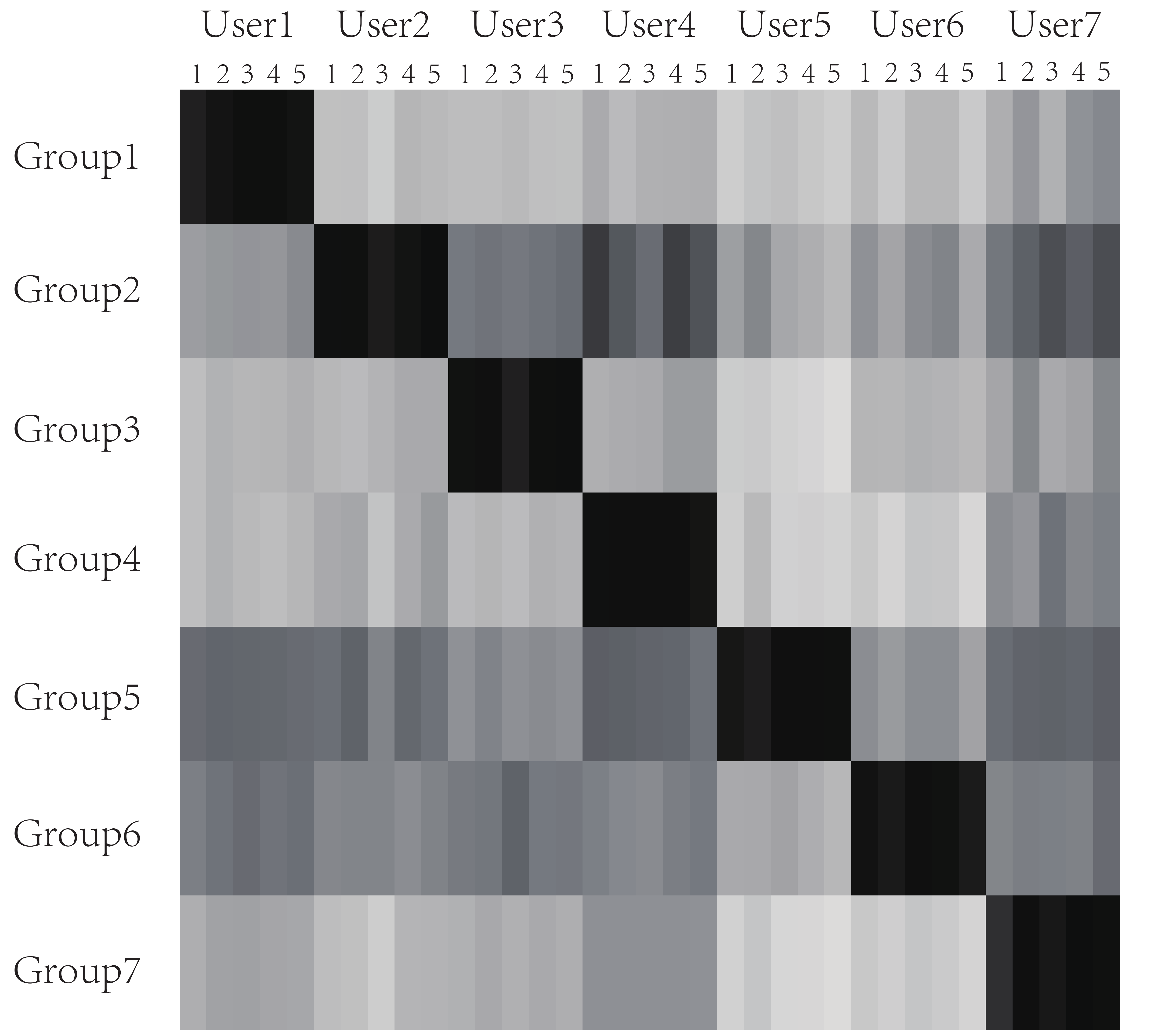}
\caption{Self-similarity test on seven distinct users with five trails for each.} \label{fig5_3_1}
\end{figure}

The first issue is the practicability. We intend to demonstrate that there is enough discrimination between authorized users and unauthorized users in our system. Here is a self-similarity test shown in Fig~\ref{fig5_3_1}. We calculated TSS between the password of each of 7 distinct users, where each training data group contains 5 samples. Setting $\mu^{(1)}=\mu^{(2)}=\cdots=\mu^{(6)}=1/6$, the result shows that the TSS from a sample to the group where it is belong to is much higher than that of other groups. Self-similarity within each group fits well with our expectation.

The next step is to show the discrimination in practical usage. We selected 15 user's writing data and trained each user's system with 5 trails. Under the threshold $\delta=0.55$, the result of $FNR$ (false negative rate) and $FPR$ (false positive rate) is shown in Fig~\ref{fig5_3_3}(left), from which we got the average $FNR=1.78\%$ and $FPR=6.70\%$. So our system performs high self and non-self discrimination under proper threshold.

\subsection{Robustness to Mimic Attacks} \label{Attack}
\begin{figure}[htbp]
\centering
\includegraphics[width=3in]{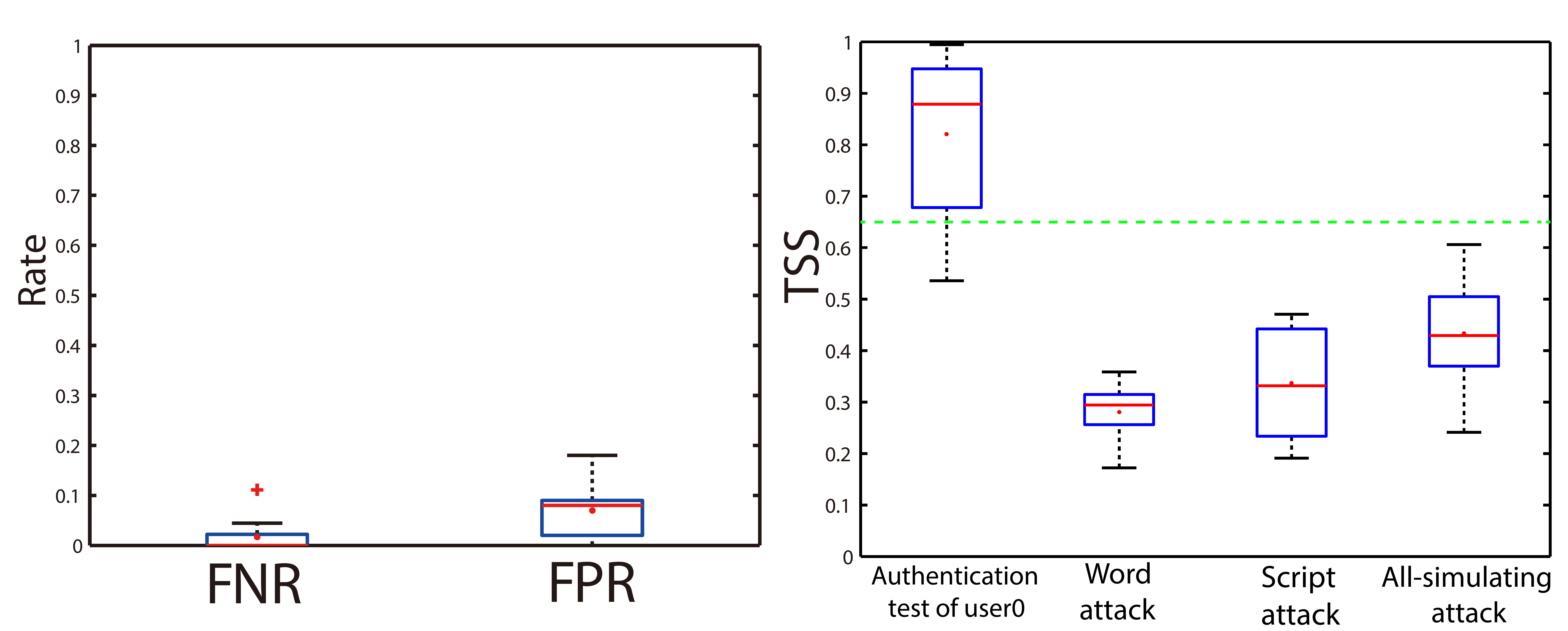}
\caption{$FNR$ and $FPR$ of test (left). and The $TSS$ of three types of attack (right).} \label{fig5_3_3}
\end{figure}

Generally, the resistance to various attack is a vital issue to an authentication system. The traditional authentication methods such as PIN
code and fingerprint identification, can be attacked in some ways.
For instance, if the user is recorded by a video while typing PIN
code, the password is easily to be stolen; another research \cite{Attack1} showed that if someone is photographed while waving hand, his fingerprint can be restored by attacker only using the photo. So our
system should have resistance to these attacks. We trained the
system by 25 "love"s written by a user and tested 3 types of attack
methods. The result is shown in Fig~\ref{fig5_3_3}(right).

\subsubsection{Word attack}
One of the simplest attack method is simulated by letting the
attacker only knows the word the user using. We asked 15
attackers to write word "love" by their habit, for 10 trials each.
\subsubsection{Script attack}
This happens if the attacker gets the script the user
has written. For example, the user wrote the word on a piece of
paper for authentication and the attacker got that paper. 15 attackers are asked to forge the script, for 10 trials each.
\subsubsection{All-simulating attack}
\begin{figure}[htbp]
\centering
\includegraphics[width=3in]{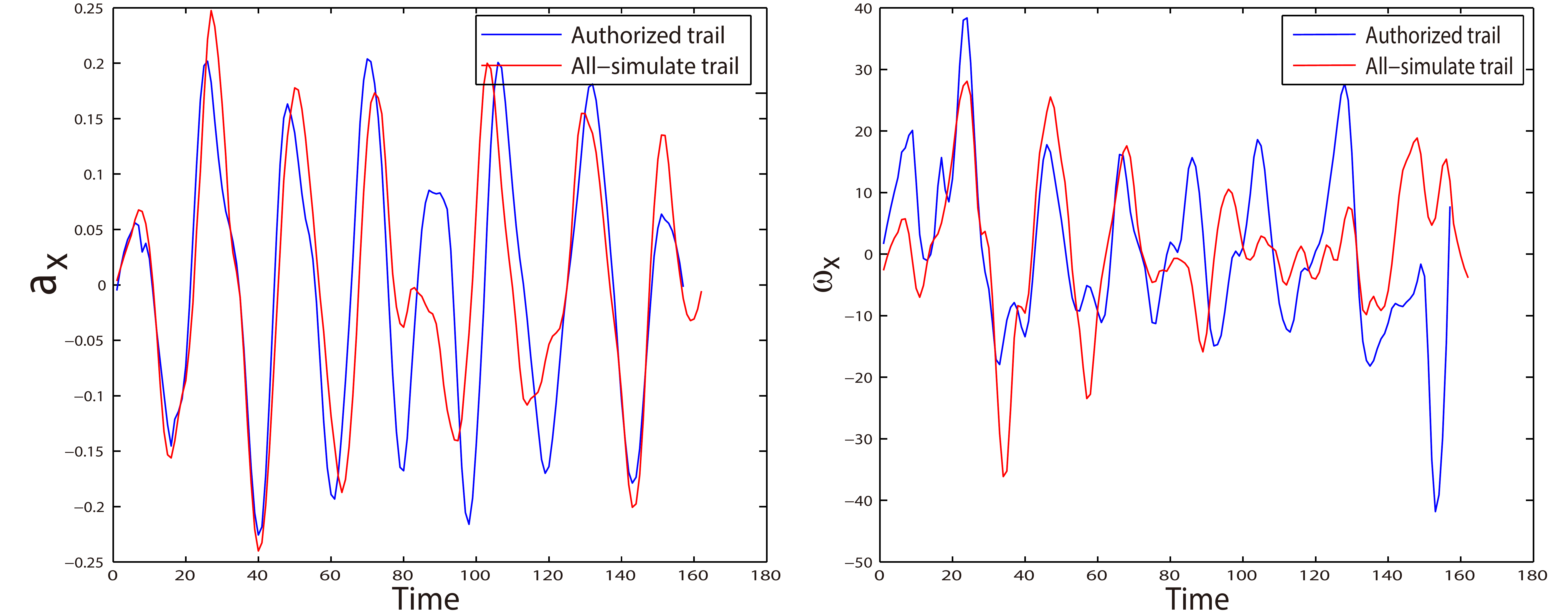}
\caption{Comparison between all-simulating attack and authorized
trail, $a_x$ (left) and $\omega_x$ (right).} \label{fig5_attack3}
\end{figure}

If the attacker records a video of the user's wrist movement while the user is
doing authentication, the attacker might be able to simulate all the
writing process and finally pass the authentication. This is the
most threatening and challenging attack. We recorded the video while
the user was writing password, and then let 15 attackers tried their best to simulate the user's writing for 10 trials each. Fig~\ref{fig5_attack3} shows that the $a_x$ curve
is very close but the $\omega_x$ curve is distinct.

The result shows that the $TSS$ of
attacker increases since the
attacker is getting more and more information of user's authentication process. Through the attack simulation, we found that although the
attacking data can not be completely distinguished from user's test data, there is a boundary lying between the TSS of attackers and that of the authorized users. Taken together security and convenience, the threshold is suggested to set as $\delta=0.65$ although the $TPR$ (True Positive Rate)
decreases to $82\%$. This means that in order to defend attacks better, the $TPR$ is sacrificed a little bit. Fortunately, we came up with a practical method to increase the $TPR$ introduced in Section~\ref{ROC}.

\subsection{Fault Tolerance to Training Data}\label{FaultTolerance}
The authentication system is supposed to be well trained by the user before being put into service. However, the
user might provide some unreasonable training data: words written in abnormal ways. This might happen for example when the user provides the training data in hurry, or when he is interrupted by others. So it is necessary to check the overall performance of the system when there are
improper data mixed in the training data group, namely, the fault tolerance.

\begin{figure}[htbp]
\centering
\includegraphics[width=2.3in]{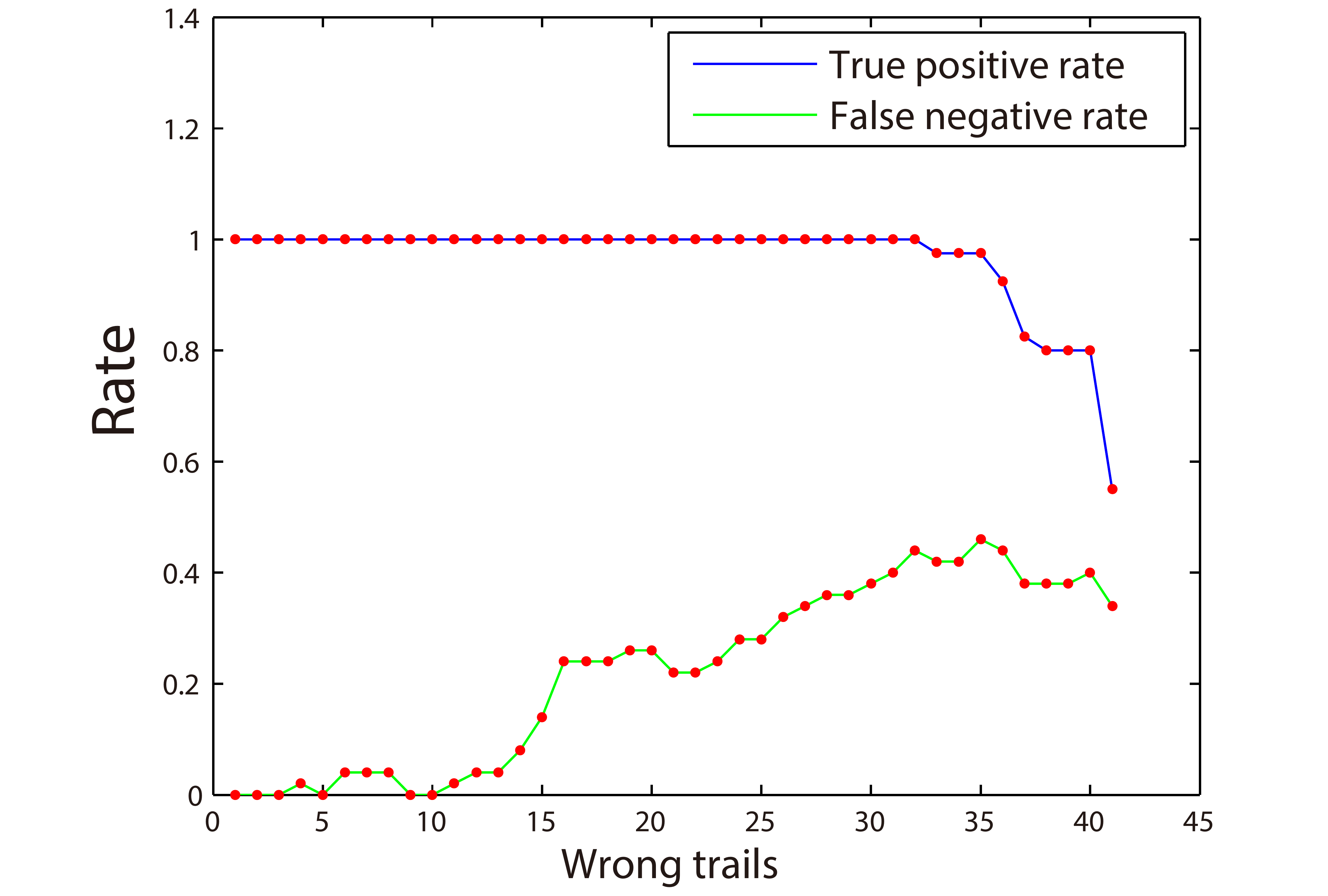}
\caption{Fault tolerance test.} \label{fig5_fault}
\end{figure}

We trained the system by 10 original trails and added wrong trails to
training groups little by little. We chose 50 fine trails and 50 bad
trails for testing. The results of $TPR$ and $FNR$ are shown in
Fig~\ref{fig5_fault}. We can see that the $TPR$ is always equal to 1
at the beginning, because our algorithm gives more weights to
more similar trails when calculating the DTW distance from a testing data to a group. The $FNR$ increases as the rate of bad data
raises, because the group ideal DTW distance cannot provide the
relative distance we want. Fortunately, we could control the $FNR$
under $5\%$ if the percentage of wrong data is below 50\%.
It's a fine result, indicating that our system is robust towards abnormal training data.

\subsection{Discrimination of Different Motion Dimensions} \label{ROC}
\begin{figure}[htbp]
\centering
\includegraphics[width=3in]{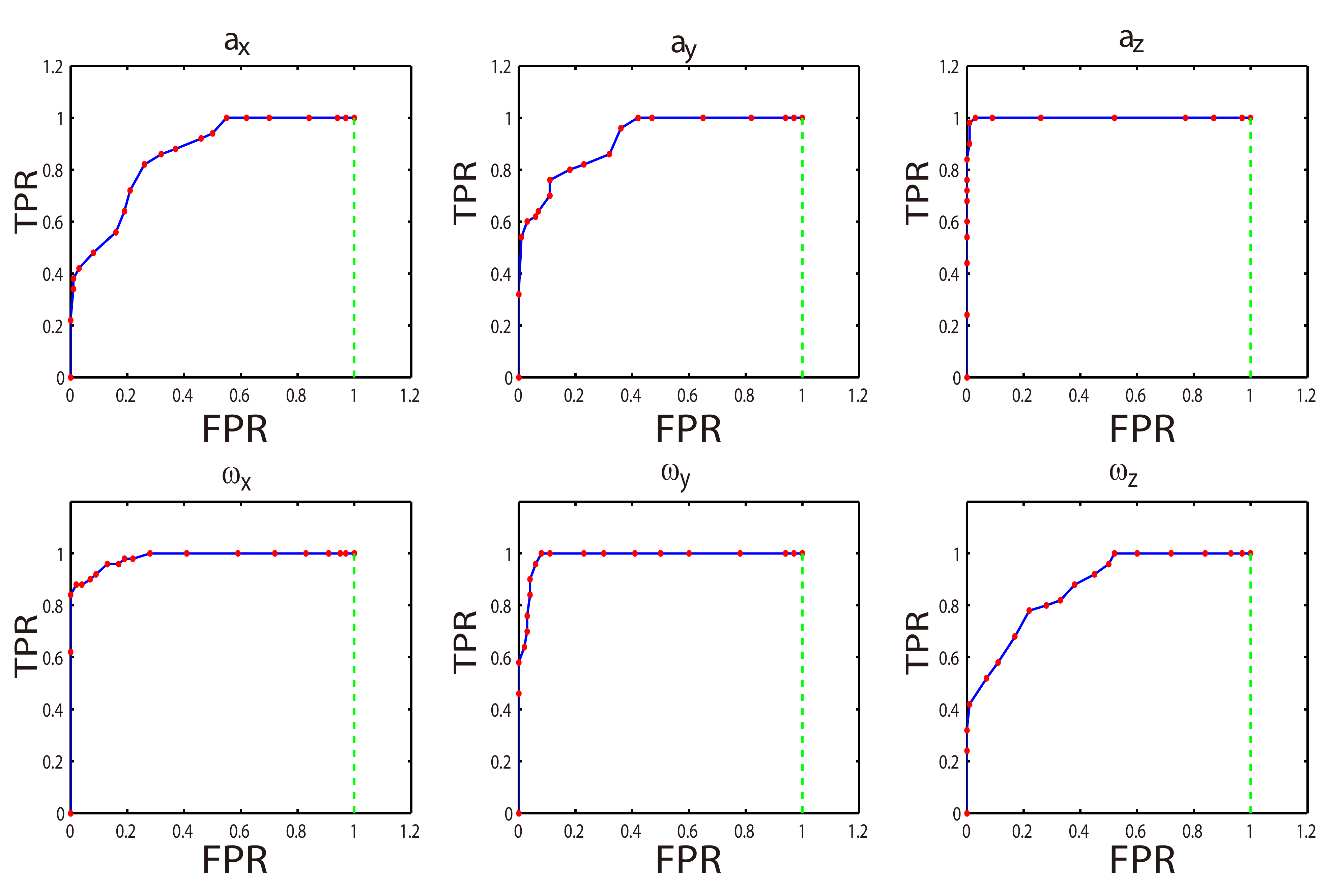}
\caption{The AUC test of each motion dimension. } \label{fig5_AUC}
\end{figure}

In order to find out the discrimination of 6 motion dimensions
$a_x,~a_y,~a_z,~\omega_x,~\omega_y,~\omega_z$, the AUC (Area Under Curve) test is implemented for each motion dimension basing on its ROC curve, the results are shown in Fig~\ref{fig5_AUC}. AUC value for each  motion dimension is:
$(A_1,A_2,\cdots,A_6)=(0.8556,~0.9130,~0.9985,~0.9839,~0.9851,~0.8682)$ and the total $AUC$ is 0.947.
The discrimination of an motion dimension is better if the AUC value is
higher, so we distribute the weight $\boldsymbol{\mu}$ based on
AUC value.
We want the performance of our system to satisfy at least $AUC>0.85$ for each motion dimension. Therefore, we distribute $\boldsymbol{\mu}$ as:
\[
B_i=\min(A_i-0.85,0),~
\mu^{(i)}=\frac{B_i}{\sum_{i=1}^6B_i}.
\]
And we got
\[\boldsymbol{\mu}=(0.0111,~0.1249,~0.2945,~0.2655,~0.2679,~0.0361).\]
Then we did the attack simulation in Section~\ref{Attack} again.  The training set is composed of 10 samples from one person and another 100 samples from 20 different persons, and the test set is composed of 40 samples from the first person. The result shows that the $FPR$ of all attacks is zero but the $TPR$ increases up to 90\% when we still set threshold $\delta=0.65$. Even if with threshold of $\delta=0.62$, we still can block all the attacks, with $TPR$ increasing to 94\%. Also the $AUC$ value of the whole system increased to 0.983.

\subsection{Contrast with SVM}\label{SVM}
The support vector machine (SVM) has been widely used in behavior-based authentication approaches, and sometimes could achieve good results. Therefore, we further contrast our approach with the SVM. We will show the results and discuss the flaw of this method. The SVM method was implemented for our authentication scenario by following steps.
\subsubsection{Feature extraction}
Each trail has 6 motion dimensions, for each column 9 features are extracted.

$\bullet$ Statistical features: Mean value, Minimum and Maximum
value, Range, Variance, Kurtosis and Skewness.

$\bullet$ Frequency-Domain features: Energy and Entropy. Let $v$ be the vector after Fast
Fourier Transformation (FFT), then $Energy(v)=\sum_{i=1}^n |v_i|^2=v^Tv$, $Entropy(v)=\sum_{i=1}^n |v_i|^2\ln|v_i|^2.$

$\bullet$ Additional features: We added two vectors $Peak$ and $Dis$ as features.
For $N(N=1000)$ times we randomly chose 2 points in each motion dimension and calculated their difference. The we got an empirical probability distribution of the difference of points in each motion dimension, called $Dis$. Another feature, $Peak$, was the peak values of the trails.

Thus, there are totally 54 features plus 2 additional vectors of features. For the first 54 features, they may be strongly related. Since the feature dimension is high, we use $Lasso$ ($\ell_1$ penalty) regression to extract important features.

\subsubsection{Feature selection}
\begin{figure}[htbp]
\centering
  \includegraphics[width=0.4\textwidth]{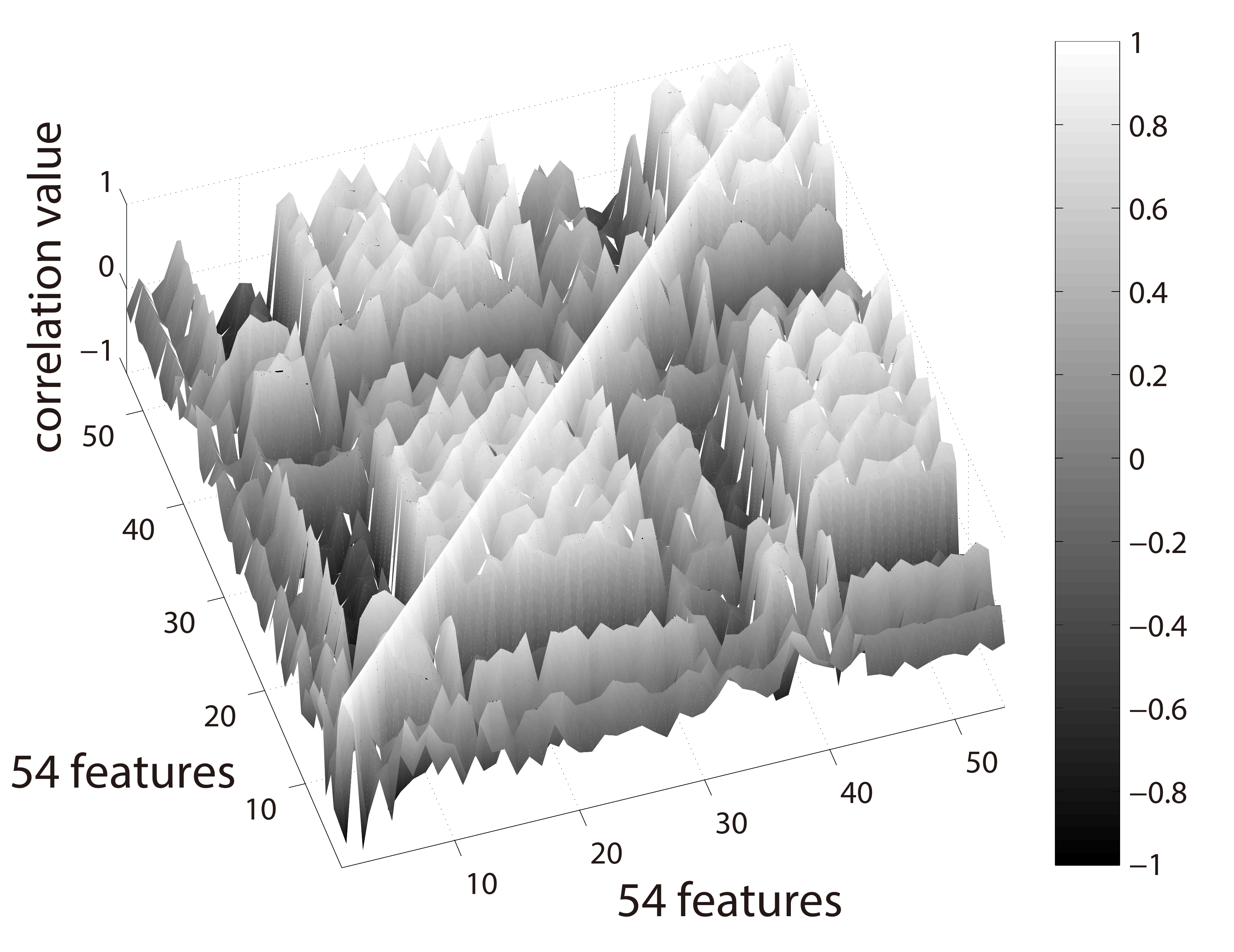}\\
  \caption{Correlation matrix of 54 features.}
  \label{5_1}
\end{figure}

For the first 54 features, they may be strongly related. In order to
reduce time expense, we need to find important and unrelated
features. Fig~\ref{5_1} shows the correlation matrix of them.

\begin{figure}[htbp]
\centering
  \includegraphics[width=0.22\textwidth]{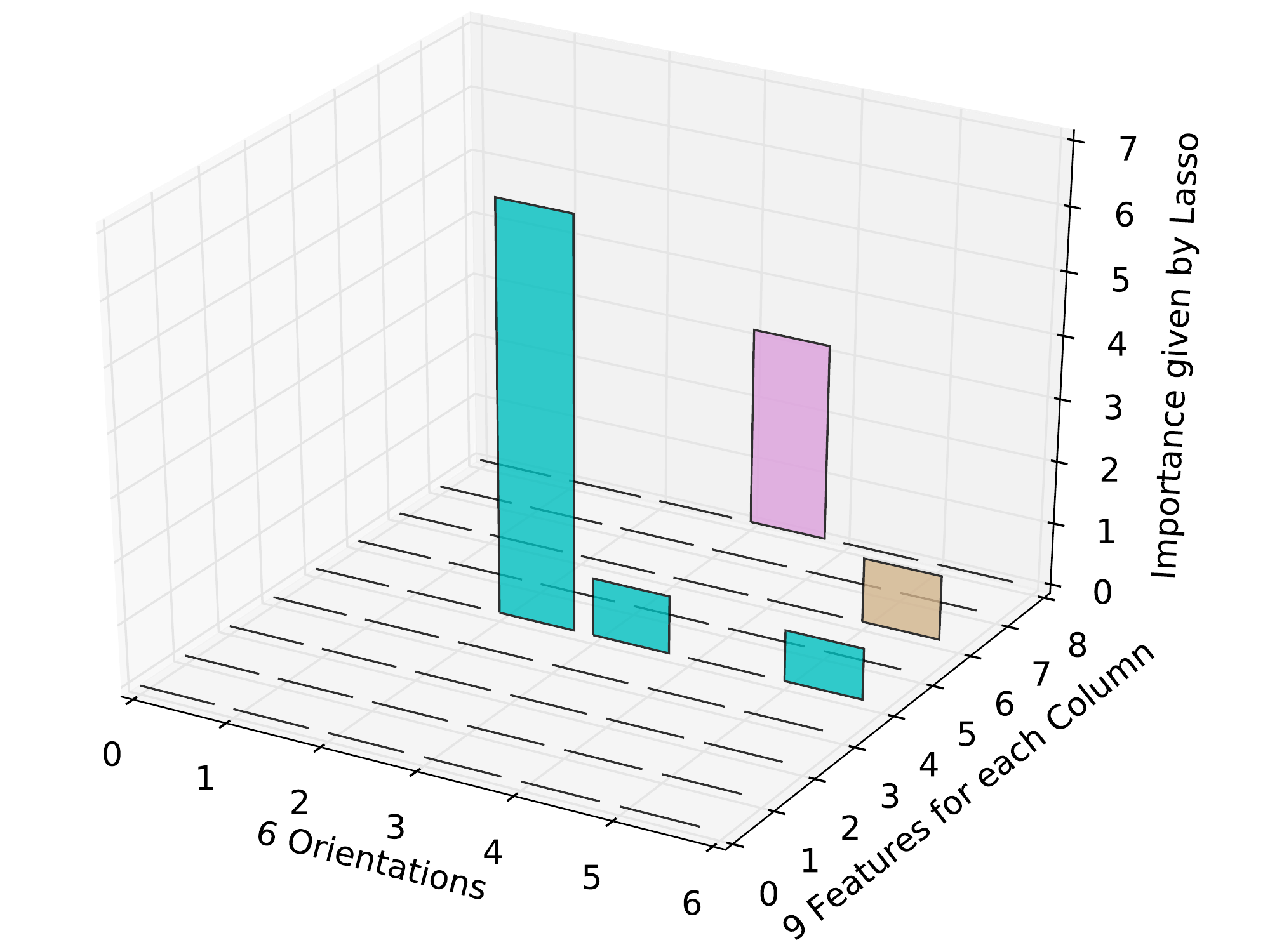}
  \includegraphics[width=0.22\textwidth]{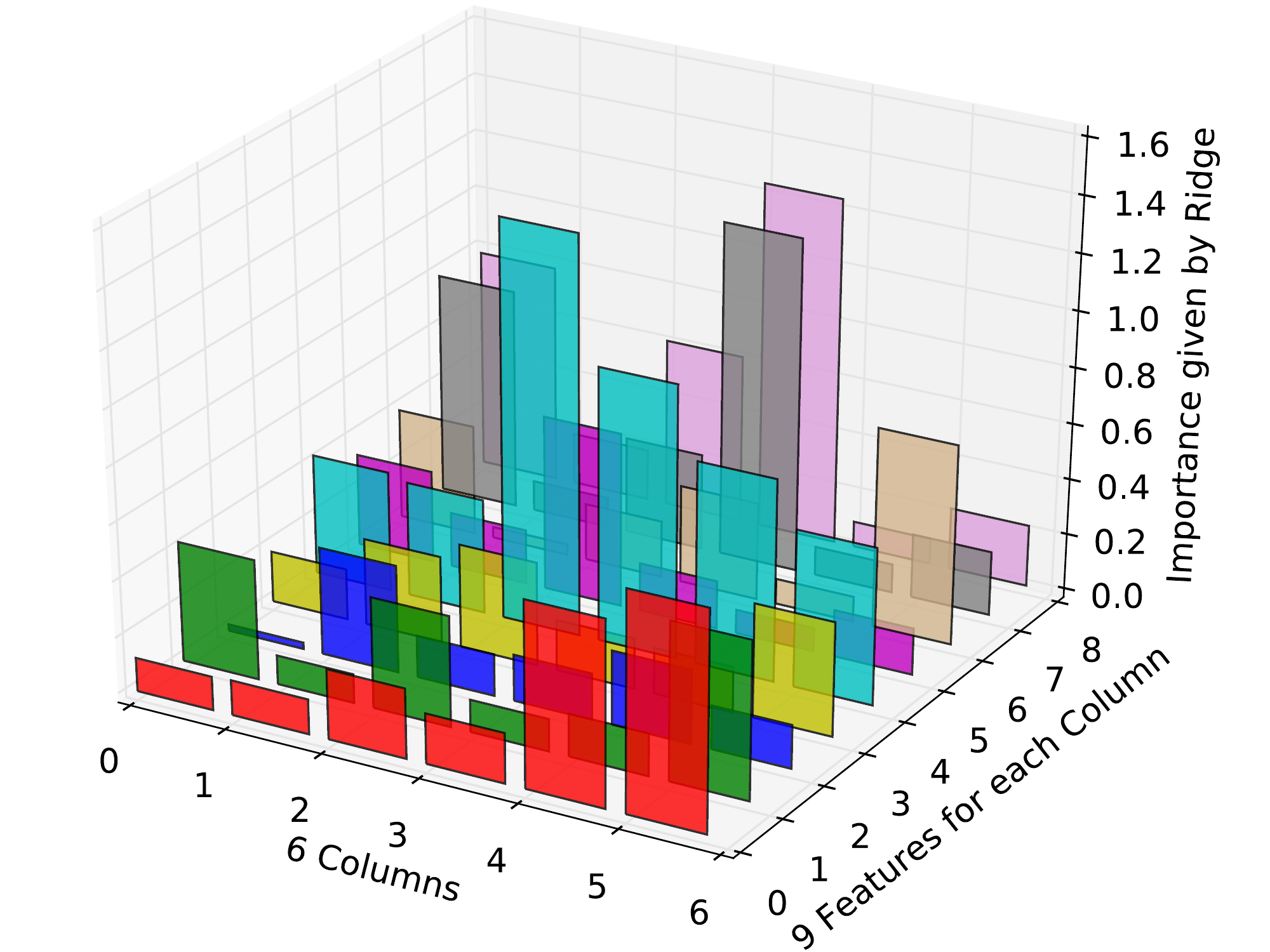}
  \caption{$\beta$ given by $Lasso$(left) and $Ridge$(right)}
  \label{5_2}
\end{figure}

The regularization method provides a very good way to estimate the
contribution of each feature. Two popular regularization methods are
$\ell_1(Lasso)$ and $\ell_2(Ridge)$ regression. The linear method to find
parameter $\beta$ is to directly minimize $\|X\beta-Y\|_2^2$, while
the regularization method just adds one more item:
\[\ell_1:\ \min_{\beta}\|X\beta-Y\|_2^2+\lambda\|\beta\|_1^2\]
\[\ell_2:\ \min_{\beta}\|X\beta-Y\|_2^2+\lambda\|\beta\|_2^2\]
where $\lambda$ is a given parameter. $\beta$ given by $\ell_2$
regression shows the contribution of each feature(shown in
Fig~\ref{5_2}  left); while $\beta$ given by $\ell_1$ regression tends
to be sparse(shown in Fig~\ref{5_2} right): if two important
features are strongly related, then only one of them will be
reported important.

\subsubsection{Applying SVM}
Now we have 400 samples of 20 words with each
word written for 20 times. 5-cross validation was applied and each time the
mean precision was recorded; the final mean
average precision $MAP$ equals to the mean value of 5 recorded
precisions.
\begin{table}[htbp]
\centering
\begin{tabular}{c|cc}
  \hline
  Feature selection & 54 features & all features \\
  \hline
  None & 95.4\% & 98.8\% \\
  $Lasso$ & 87.8\% & 98.8\% \\
  \hline
\end{tabular}
\caption{$MAP$s with/without feature selection.}
\end{table}

\subsubsection{Flaw of SVM}
However, compared to our method, SVM is hard to detect abnormal points, in that provided a testing data, SVM would classify it into one existed class, even if the input is not supposed to be classified into any of the existed group. This flaw makes it fairly unreliable.

We collected 10 samples for each of the 10 different
words (not "love"), and 20 samples for one additional word "love",
which is the password. Then we choose some training data in each group and the rest are testing data. Use SVM to train the training data and then classify the testing data, then 5 of them (like word "book") are classified as "love",
which makes the authentication extremely insecure!


\section{Conclusion}
In this paper, we present a new approach to user authentication using the behavioral biometrics provided by wrist-worn devices. Our approach focuses on effective modeling methods to obtain fine-grained writing-interaction metrics, which have two advantages over other metrics. First, fine-grained writing-interaction metrics can distinguish a user accurately with very few strokes. Second, the metrics are hard to attack even if the whole authentication process are recorded in attackers' video tape. Extensive experimental results show that the proposed approach can identify users in a convenient and accurate manner, making itself suitable for online and mobile authentication.


\begin{thebibliography}{}

\bibitem{Password1}
Klein D V. \textit{Foiling the cracker: A survey of, and improvements to,
password security}, in Proceedings of the 2nd USENIX Security
Workshop. 1990: 5-14.

\bibitem{Password2}
Ives B, Walsh K R, Schneider H. \textit{The domino effect of password
reuse}, in Communications of the ACM, 2004, 47(4): 75-78.

\bibitem{Password3}
Ur B, Bees J, Segreti S M, $et~al$. \textit{Do Users' Perceptions of Password
Security Match Reality?} in Proceedings of the SIGCHI Conference on
Human Factors in Computing Systems (CHI). 2016.

\bibitem{Fingerprint1}
Jain A K, Hong L, Pankanti S, $et~al$. \textit{An identity-authentication
system using fingerprints}, in Proceedings of the IEEE, 1997, 85(9):
1365-1388.

\bibitem{Iris1}
Chong S C, Teoh A B J, Ngo D C L. \textit{Iris authentication using
privatized advanced correlation filter}, inInternational Conference
on Biometrics. Springer Berlin Heidelberg, 2006: 382-388.

\bibitem{Accelerometer1}
Liu J, Zhong L, Wickramasuriya J, $et~al$. \textit{User evaluation of
lightweight user authentication with a single tri-axis
accelerometer}, in Proceedings of the 11th International Conference
on Human-Computer Interaction with Mobile Devices and Services. ACM,
2009: 15.

\bibitem{Accelerometer2}
Yang L, Guo Y, Ding X, $et~al$. \textit{Unlocking Smart Phone through
Handwaving Biometrics}, in IEEE Transactions on Mobile Computing,
2015, 14(5): 1044-1055.

\bibitem{Wrist1}
Yang J, Li Y, Xie M. \textit{MotionAuth: Motion-based authentication for
wrist worn smart devices}, in Pervasive Computing and Communication
Workshops (PerCom Workshops), 2015 IEEE International Conference on.
IEEE, 2015: 550-555.

\bibitem{Wrist2}
Mare S, Markham A M, Cornelius C, $et~al$. \textit{ZEBRA: Zero-Effort
Bilateral Recurring Authentication}, in IEEE symposium on security
and privacy, 2014: 705-720.

\bibitem{Attack1}
\textit{Fingerprint Biometrics Hacked Again.}\\
\url{http://www.ccc.de/en/updates/2014/ursel}.

\bibitem{Biometrie}
Ohtsuki T, Kamoi H. \textit{Biometrie authentication using hand movement information from wrist-worn PPG sensors}, in Personal, Indoor, and Mobile Radio Communications (PIMRC), 2016 IEEE 27th Annual International Symposium on. IEEE, 2016: 1-5.

\bibitem{Signature}
Nassi B, Levy A, Elovici Y, et al. \textit{Handwritten Signature Verification Using Hand-Worn Devices}. arXiv preprint arXiv:1612.06305, 2016.

\bibitem{WristSense1}
Cole R J, Kripke D F, Gruen W, et al. \textit{Automatic sleep/wake
identification from wrist activity}, in Sleep, 1992, 15(5): 461-469.

\bibitem{WristSense2}
Dong Y, Scisco J, Wilson M, et al. \textit{Detecting periods of eating
during free-living by tracking wrist motion}, in IEEE journal of
biomedical and health informatics, 2014, 18(4): 1253-1260.

\bibitem{SGfilter}
Savitzky, A., Golay, M.J.E. \textit{Smoothing and Differentiation of Data by Simplified Least Squares Procedures}, in Analytical Chemistry, 1964. 36(8): 1627¨C39.


\bibitem{DTWpaper}
Keogh E, Pazzani M J. \textit{Derivative Dynamic Time Warping}, in siam international conference on data mining, 2001: 1-11.

\bibitem{DTWtoolbox}
Roger Jang. \textit{Machine Learning Toolbox.}\\
\url{http://www.mirlab.org/jang/matlab/toolbox/ machineLearning/}, 05/15, 2011.

\bibitem{SDK}
\textit{Microsoft band SDK documentation.}\\
\url{https://developer.microsoftband.com/Content/docs/Microsoft \%20Band\%20SDK.pdf}.


\end{thebibliography}
\end{document}